\documentclass{article}
\pdfoutput=1

\usepackage{enumerate}
\usepackage{amsfonts}
\usepackage{amsmath}
\usepackage{comment}
\usepackage{textcomp}
\usepackage{amsthm, tikz}
\usepackage{amssymb}
\usepackage{color}
\usepackage{graphicx}
\usepackage{bbm}
\usepackage[matrix,arrow,curve]{xy}
\usepackage{array}
\usepackage{mathtools}

\def\be{\begin{eqnarray}}
\def\ee{\end{eqnarray}}
\def\nn{\nonumber}

\def\Tr{{\rm Tr}\,}

\def\l[{\phantom.[}

\def\vac{0}

\hoffset= - 1.0in         % switch off for draft style
\begin{document}
\title{{\bf {Towards effective topological field theory for knots
}\vspace{.2cm}}
\author{{\bf A. Mironov$^{a,b,c,d}$}\ and \ {\bf A. Morozov$^{b,c,d}$}}
\date{ }
}

\maketitle

\vspace{-5.5cm}

\begin{center}
%\hfill FIAN/TD-05/15\\
\hfill IITP/TH-05/15
\end{center}

\vspace{4.2cm}

\begin{center}

$^a$ {\small {\it Lebedev Physics Institute, Moscow 119991, Russia}}\\
$^b$ {\small {\it ITEP, Moscow 117218, Russia}}\\
$^c$ {\small {\it National Research Nuclear University MEPhI, Moscow 115409, Russia }}\\
$^d$ {\small {\it Institute for Information Transmission Problems, Moscow 127994, Russia}}\\
\end{center}

\vspace{1cm}

\begin{abstract}
Construction  of (colored) knot polynomials for double-fat graphs is further generalized
to the case when "fingers" and "propagators" are substituting ${\cal R}$-matrices
in arbitrary closed braids with $m$-strands.
Original version of \cite{MMMRS} corresponds to the case $m=2$,
and our generalizations sheds additional light on the structure of those mysterious
formulas.
Explicit expressions are now combined from Racah matrices of the type $R\otimes R\otimes\bar R
\longrightarrow \bar R$ and mixing matrices in the sectors $R^{\otimes 3}\longrightarrow Q$.
Further extension is provided by composition rules, allowing to glue two blocks,
connected by an $m$-strand braid
(they generalize the product formula for ordinary composite knots with $m=1$).
\end{abstract}

\vspace{1cm}

\section{Introduction}

Reshetikhin-Turaev (RT) formalism \cite{RT} remains the most effective approach for
actual evaluation of colored HOMFLY polynomials \cite{Knotpols}
\be
H_R^{\cal L} =  \left< {\rm Tr}_R\ {\rm Pexp}\left(\oint_{\cal L} {\cal A}\right)\right>^{\rm CS}
\ee
Derivation of the RT rules from functional integral \cite{WitJones,Kaul,Labastida},
is only partly understood, see \cite{MorSmi,DSS} for recent comments.
A partly alternative approach is implied by Khovanov calculus \cite{Kh}-\cite{Witlast},
see \cite{DM3}, but it is also related to the RT approach via Kauffman's $R$-matrix \cite{KaufR}.

Ideally, the RT formalism allows one to cut the link diagram
(an oriented graph with black and white vertices of valences $(2,2)$) into arbitrary fragments,
and then contract tensors associated with these fragments, to obtain $H_R$.
The smallest possible fragments are vertices, represented by quantum ${\cal R}$-matrices
and contractions involve additional insertions of weight matrices, one per each Seifert cycle
(alternatively one can make contractions dependent on a choice of direction in the plane).
In this (original) form the formalism depends on particular representations of the gauge group
and even the number of tensor indices depend on representation.

Representation theory allows one to switch to a dual Tanaka-Krein description,
where indices label representations themselves (Young diagrams in the case of $SU_q(N)$)
in the space of intertwining operators
rather than vectors in representation spaces.
It is this formulation, which provides conceptually and practically important
expressions for link polynomials.

This approach is absolutely universal, but to make it calculationally effective one
still needs to deal with specially selected fragments of link diagrams.
So far two standard classes were thoroughly analyzed. Both are made from braids.

One class is that of closed $m$-strand braids of \cite{MMMkn12,IMMM,anopaths}, where
one distinguishes between $m-1$ types of ${\cal R}$-matrices, acting on pairs of
adjacent braids and related by {\it mixing matrices},
\be
{\cal R}_j = {\cal U}_j{\cal R}{\cal U}_j^{-1}, \ \ \ \ j = 1,\ldots,m-1
\label{mixma}
\ee
Each link diagram is Reidemeister-equivalent to such closed braid,
which substitutes ${\cal L}$ by a sequence of integers
$(a_{11},\ldots,a_{1,m-1},a_{21},\ldots, a_{n,m-1})$ (this map is highly ambiguous),
and
\be
H_R^{\cal L} =  \sum_{Q\in R^{\otimes m}} d_Q \Tr_{\mu_Q}
\left(\prod_{i=1}^n \prod_{j=1}^{m-1} {\cal R}_j^{a_{ij}}\right)
\label{Hbraid}
\ee
where $R^{\otimes m} = \sum_Q W_Q\otimes Q$ is decomposed into a sum of the Young diagrams $Q$ of size $|Q|=m|R|$.
The coefficients $d_Q$ are the corresponding quantum dimensions of the $SU_q(N)$ representations
and vector spaces  $W_Q$ of intertwining operators have dimensions $\mu_Q$
(which depend also on $R$ and $m$).
The properly defined quantum ${\cal R}$-matrix acting on the product $R_1\otimes R_2$ acts diagonally in irreducible components $Q$
of its decomposition with the eigenvalues $q^{\varkappa_Q}$, where $\varkappa_Q$ is the eigenvalue of the Casimir operator.
Thus ${\cal R}_j$ and ${\cal U}_j$ can be considered as matrices acting on the spaces $W_Q$
and this is what we call their Tanaka-Krein realization.
This story is well familiar from the Rosso-Jones formula \cite{RJ}
for the torus knots/links and is related to the theory of cut-and-join operators \cite{MMN1}.
For links up to $\#_{\rm link\ comps}$ different representations appear in this formula.
Also, if some strands in the braid go in the opposite direction, they carry
the conjugate representation $\bar R$ and the mixing matrices ${\cal U}_i$
change appropriately.
Mixing matrices are contractions of the Racah matrices ($6j$-symbols) \cite{Racah}
and are rather difficult to calculate. This, together with insufficient computer
power, is the main obstacle against using
(\ref{Hbraid}) for practical calculations of colored HOMFLY polynomials for complicated knots
(at present already $R=[21]$ at $m=4$ strands and $R=[31]$ at $m=3$,
i.e. $12$ strands in the cabling method of \cite{anopaths,anoMcabling},
are nearly unaffordable, straightforwardly available are $R=[21]$ at $m=3$ \cite{Ano21,MMM21}).

Another class is formed by the 2-bridge knots.
These are made from $4$-strand braids with two strands pointing in one and two in the other
directions.
These are not the closed braids and they are distinguished by a relation to $4$-point conformal blocks,
where the Racah and ${\cal R}$-matrices, or $S$ and $T$ accordingly play the role of modular transformations.
This interpretation \cite{WitJones,Kaul} allows one to associate with braids not only traces,
but also the matrix elements,
in particular knot polynomials for the $2$-bridge knots are provided
by "vacuum averages".
The restriction to the {\it two}-bridge knots involves only the Racah matrices

\be
\begin{picture}(300,120)(0,-50)
\put(0,0){\line(1,0){50}}
\put(0,0){\line(-1,-1){20}}
\put(0,0){\line(-1,1){20}}
\put(50,0){\line(1,-1){20}}
\put(50,0){\line(1,1){20}}
\put(97,0){\vector(1,0){120}}
\put(115,20){\mbox{$S_{Y\!X}\left(\begin{array}{cc} R_2 &R_3 \\ R_1 &R_4 \end{array}\right)$}}
\put(250,-20){\line(0,1){40}}
\put(250,20){\line(-1,1){20}}
\put(250,20){\line(1,1){20}}
\put(250,-20){\line(-1,-1){20}}
\put(250,-20){\line(1,-1){20}}
\qbezier(235,40)(250,60)(265,40)
\put(237,42){\vector(-1,-1){2}}
\put(263,42){\vector(1,-1){2}}
\put(227,60){\mbox{$T_Y(R_2,R_3)$}}
\put(255,2){\mbox{$Y$}}
\put(23,5){\mbox{$X$}}
\put(-15,-25){\mbox{$R_1$}}
\put(-15,20){\mbox{$R_2$}}
\put(54,20){\mbox{$R_3$}}
\put(54,-25){\mbox{$R_4$}}
\put(220,-35){\mbox{$R_1$}}
\put(220,30){\mbox{$R_2$}}
\put(270,30){\mbox{$R_3$}}
\put(270,-35){\mbox{$R_4$}}
\end{picture}
\ee

\noindent
with all the four $R_1,R_2,R_3,R_4$ equal to either $R$ or $\bar R$, which we call simply $S$ or $\bar S$ depending on the direction of arrows.

\bigskip

Connecting the external double-lines of the 2-bridge building blocks

\begin{picture}(300,125)(-100,-30)
\put(0,0){\line(1,0){30}}
\put(0,0){\line(0,1){60}}
\put(0,60){\line(1,0){30}}
\put(30,0){\line(0,1){60}}
\put(2,2){\line(1,0){26}}
\put(2,2){\line(0,1){56}}
\put(2,58){\line(1,0){26}}
\put(28,2){\line(0,1){56}}
\put(5,0){\line(-1,-2){9.5}}
\put(8,0){\line(-1,-2){10}}
\put(25,0){\line(1,-2){9.5}}
\put(22,0){\line(1,-2){10}}
\put(5,60){\line(-1,2){9.5}}
\put(8,60){\line(-1,2){10}}
\put(25,60){\line(1,2){9.5}}
\put(22,60){\line(1,2){10.0}}
\put(-15,-22){\mbox{$X$}}
\put(38,-22){\mbox{$\bar X$}}
\put(-15,80){\mbox{$Y$}}
\put(38,80){\mbox{$\bar Y$}}
\put(7,28){\mbox{$B_{Y\!X}$}}
\end{picture}

\noindent
one can make entire networks.
If one considers no more than 2-"particle" irreducible graphs (the term "particle" appeals to the line in the fat graph),
the HOMFLY polynomial of the corresponding link diagrams are obtained just by summing over
the indices. Let us re-draw the graphs with straight lines denoting double-fat fingers and propagators $B_{YX}$ (i.e. 2-bridge building blocks, or 4 strand braids), and circles consisting of
{\it two}-strand braids (i.e. "particles") connecting $B_{YX}$. Then, the typical diagrams are

\be\label{fig}
\begin{picture}(300,180)(-80,-80)
\put(-100,80){\mbox{forbidden:}}
\put(-80,-30){\circle{40}}
\put(-140,30){\circle{40}}
\put(-20,30){\circle{40}}
\put(-120,30){\line(1,0){80}}
\put(-94,-16){\line(-1,1){32}}
\put(-66,-16){\line(1,1){32}}
\put(180,80){\mbox{allowed:}}
\put(200,0){\circle{40}}
\put(120,0){\circle{40}}
\put(280,0){\circle{40}}
\put(140,0){\line(1,0){40}}
\put(220,0){\line(1,0){40}}
\put(214,-14){\line(1,-1){20}}
\put(186,-14){\line(-1,-1){20}}
\put(186,14){\line(-1,1){20}}
\put(214,14){\line(1,1){20}}
\put(106,-14){\line(-1,-1){20}}
\put(106,14){\line(-1,1){20}}
\put(294,-14){\line(1,-1){20}}
\put(294,14){\line(1,1){20}}
\put(118,-2){\mbox{$Y$}}
\put(198,-2){\mbox{$X$}}
\put(278,-2){\mbox{$Z$}}
\put(80,-2){\mbox{$\ldots$}}
\put(310,-2){\mbox{$\ldots$}}
\put(196,-35){\mbox{$\ldots$}}
\put(196,30){\mbox{$\ldots$}}
\put(92,-35){\mbox{$B_{0Y}^{(1)}$}}
\put(92,30){\mbox{$B_{0Y}^{(m)}$}}
\put(286,30){\mbox{$B_{0Z}^{(1)}$}}
\put(286,-35){\mbox{$B_{0Z}^{(n)}$}}
\put(238,-40){\mbox{$B_{0X}^{(1)}$}}
\put(143,-40){\mbox{$B_{0X}^{(k)}$}}
\put(140,35){\mbox{$B_{0X}^{(k+2)}$}}
\put(238,35){\mbox{$B_{0X}^{(l)}$}}
\put(145,5){\mbox{$B_{YX}^{(k+1)}$}}
\put(227,5){\mbox{$B_{ZX}^{(l+1)}$}}
\end{picture}
\ee

\noindent
The shown allowed configuration provides
\be
H = \sum_{X,Y,Z} d_Xd_Yd_Z\cdot \Tr_{\mu_Y}\Tr_{\mu_Z}\left\{
\Tr_{\mu_X}\!\!\left(\prod_{\alpha=1}^{k} B^{(\alpha)}_{\vac X} \cdot B^{(k+1)}_{Y\!X}\cdot\!\!\!
\prod_{\beta=k+2}^l\!\!B^{(\beta)}_{\vac X} \cdot B^{(l+1)}_{Z\!X}\right)\right\}
\cdot \left(\prod_{\gamma=1}^mB^{(\gamma)}_{\vac Y}\right)
\cdot \left(\prod_{\delta=1}^nB^{(\delta)}_{\vac Z}\right)
\label{noloops}
\ee
Formulas of this type, for what was called {\it double-fat trees} in \cite{MMMRS},
were widely analyzed in \cite{inds,gmmms} and in the recent \cite{MMMRS}
they were successfully applied to evaluation of $[21]$-colored HOMFLY of most of knots
from the Rolfsen table \cite{katlas} and of numerous mutants, see also \cite{NRS}.

Whenever a knot possesses the both representations, as a closed braid with $m\leq 3$ and
as a double-fat tree, the corresponding HOMFLY polynomials coincide.

\bigskip

The purpose of the present note is to marry up the two above classes.
Namely, we consider a closed $m$-strand braid, cut a pair of adjacent strands
at any place and insert a 2-bridge finger $B$.
If it was just a crossing, one would insert ${\cal R}_j= {\cal U}_j{\cal R}{\cal U}_j^{-1}$.
In the case of finger $B$, one inserts ${\cal U}_jB{\cal U}_j^{-1}$.

Moreover, the finger can actually be open on the other side, which can be inserted
into another closed braid. In this way we obtain a double-fat tree, made from propagators ${\cal B}$,
which are now connected by multi-strand braids.
In other words, straight lines in allowed configuration above are still
the double-fat propagators and fingers, but circles can be braids with arbitrary number of strands,
moreover, this number can be different for different circles.
This provides an amusing set of link diagrams.

Also, if the closed braids are all $4$-strand, like the interiors of $B$,
one gets a kind of description of {\it generic} double-fat {\it graphs},
not obligatory {\it trees}.
This would complete formulation of a peculiar  {\bf new topological theory},
describing this class of link diagrams, for which only the tree
approximation was introduced and studied in \cite{MMMRS}.
This theory does not look at all like original Chern-Simons theory (CST)
which is itself topological.
Thus, one can even expect some type of {\it a duality} between CST and
this new theory.

\section{Explicit formulas}
\subsection{The case of $m=2$}

This is the setting, already considered in \cite{inds,gmmms} and \cite{MMMRS}.
In the simplest case when there is just one closed $m=2$-strand braid,
with $g+1$ pretzel fingers attached to it,

\begin{picture}(500,120)(20,-40)
\put(0,0){\line(1,0){30}}
\put(0,0){\line(0,1){60}}
\put(0,60){\line(1,0){30}}
\put(30,0){\line(0,1){60}}
\put(2,2){\line(1,0){26}}
\put(2,2){\line(0,1){56}}
\put(2,58){\line(1,0){26}}
\put(28,2){\line(0,1){56}}
\put(100,0){\line(1,0){30}}
\put(100,0){\line(0,1){60}}
\put(100,60){\line(1,0){30}}
\put(130,0){\line(0,1){60}}
\put(102,2){\line(1,0){26}}
\put(102,2){\line(0,1){56}}
\put(102,58){\line(1,0){26}}
\put(128,2){\line(0,1){56}}
\put(185,20){\mbox{$\ldots$}}
\put(250,0){\line(1,0){30}}
\put(250,0){\line(0,1){60}}
\put(250,60){\line(1,0){30}}
\put(280,0){\line(0,1){60}}
\put(252,2){\line(1,0){26}}
\put(252,2){\line(0,1){56}}
\put(252,58){\line(1,0){26}}
\put(278,2){\line(0,1){56}}
\qbezier(25,0)(25,-15)(65,-15)
\qbezier(105,0)(105,-15)(65,-15)
\qbezier(22,0)(22,-18)(65,-18)
\qbezier(108,0)(108,-18)(65,-18)
\qbezier(125,0)(125,-15)(165,-15)
\qbezier(255,0)(255,-15)(215,-15)
\qbezier(122,0)(122,-18)(165,-18)
\qbezier(258,0)(258,-18)(215,-18)
\qbezier(5,0)(5,-30)(60,-30)\qbezier(60,-30)(100,-30)(165,-30)
\qbezier(275,0)(275,-30)(215,-30)
\qbezier(8,0)(8,-27)(60,-27)\qbezier(60,-27)(100,-27)(165,-27)
\qbezier(272,0)(272,-27)(215,-27)
%
%\put(5,0){\line(-1,-2){9.5}}
%\put(8,0){\line(-1,-2){10}}
%\put(25,0){\line(1,-2){9.5}}
%\put(22,0){\line(1,-2){10}}
%\put(5,60){\line(-1,2){9.5}}
%\put(8,60){\line(-1,2){10}}
%\put(25,60){\line(1,2){9.5}}
%\put(22,60){\line(1,2){10.0}}
\put(63,-10){\mbox{$X$}}
\put(160,-10){\mbox{$X$}}
\put(220,-10){\mbox{$X$}}
\put(0,-27){\mbox{$X$}}
%\put(-15,80){\mbox{$Y$}}
%\put(38,80){\mbox{$\bar Y$}}
\put(7,28){\mbox{$B_{\vac X}^{(0)}$}}
\put(107,28){\mbox{$B_{\vac X}^{(1)}$}}
\put(257,28){\mbox{$B_{\vac X}^{(g)}$}}
\put(305,25){\mbox{{\footnotesize $B^{(p)}=P^{(n_p)}$}}}
\put(305,15){\vector(1,0){50}}
\put(380,0){\line(1,0){20}}
\put(380,0){\line(0,1){50}}
\put(380,50){\line(1,0){20}}
\put(400,0){\line(0,1){50}}
\put(420,0){\line(1,0){20}}
\put(420,0){\line(0,1){50}}
\put(420,50){\line(1,0){20}}
\put(440,0){\line(0,1){50}}
\put(455,10){\mbox{$\ldots$}}
\put(480,0){\line(1,0){20}}
\put(480,0){\line(0,1){50}}
\put(480,50){\line(1,0){20}}
\put(500,0){\line(0,1){50}}
\qbezier(395,0)(395,-10)(410,-10)\qbezier(425,0)(425,-10)(410,-10)
\qbezier(395,50)(395,60)(410,60)\qbezier(425,50)(425,60)(410,60)
\qbezier(435,0)(435,-10)(450,-10)\qbezier(485,0)(485,-10)(470,-10)
\qbezier(435,50)(435,60)(450,60)\qbezier(485,50)(485,60)(470,60)
\qbezier(385,0)(385,-20)(410,-20)\qbezier(410,-20)(430,-20)(450,-20)\qbezier(495,0)(495,-20)(470,-20)
\qbezier(385,50)(385,70)(410,70)\qbezier(410,70)(430,70)(450,70)\qbezier(495,50)(495,70)(470,70)
\put(385,23){\mbox{$n_0$}}
\put(425,23){\mbox{$n_1$}}
\put(485,23){\mbox{$n_g$}}
\end{picture}

\noindent
we get the formulas from \cite{gmmms} for the reduced HOMFLY of the pretzel knot:
\be\label{oa}
H_R^{Pr(n_0,\ldots,n_g)} =
\frac{1}{d_R}\sum_X d_X \prod_{i=0}^g \frac{(\bar S^\dagger \bar T^{n_i}S)_{\vac X}}{S_{\vac X}}
\ee
This formula is for the case of {\it odd antiparallel} \ fingers, when both
strands in the double lines are co-directed,

\begin{picture}(300,110)(-100,-30)
\put(0,0){\line(1,0){30}}
\put(0,0){\line(0,1){60}}
\put(0,60){\line(1,0){30}}
\put(30,0){\line(0,1){60}}
\put(2,2){\line(1,0){26}}
\put(2,2){\line(0,1){56}}
\put(2,58){\line(1,0){26}}
\put(28,2){\line(0,1){56}}
\put(5,0){\line(-1,-2){9.5}}
\put(8,0){\line(-1,-2){10}}
\put(25,0){\line(1,-2){9.5}}
\put(22,0){\line(1,-2){10}}
%\put(5,60){\line(-1,2){9.5}}
%\put(8,60){\line(-1,2){10}}
%\put(25,60){\line(1,2){9.5}}
%\put(22,60){\line(1,2){10.0}}
\put(-15,-22){\mbox{$X$}}
\put(38,-22){\mbox{$\bar X$}}
%\put(-15,80){\mbox{$Y$}}
%\put(38,80){\mbox{$\bar Y$}}
\put(7,28){\mbox{$P^{(n)}_{ X}$}}
\put(100,25){\mbox{=}}
\put(150,0){\line(1,0){30}}
\put(150,0){\line(0,1){60}}
\put(150,60){\line(1,0){30}}
\put(180,0){\line(0,1){60}}
\put(148,-20){\vector(1,2){10}}
\put(172,0){\vector(1,-2){10}}
\qbezier(158,60)(158,70)(148,70)\qbezier(148,70)(138,70)(138,60)
\put(138,-20){\vector(0,1){80}}
\qbezier(172,60)(172,70)(182,70)\qbezier(182,70)(192,70)(192,60)
\put(192,60){\vector(0,-1){80}}
\put(162,28){\mbox{$n$}}
\put(157,20){\mbox{${\rm odd}$}}
\put(139,-22){\mbox{$X$}}
\put(181,-22){\mbox{$X$}}
\end{picture}

\noindent
and $n$ is odd (otherwise this configuration is topologically impossible).
Note that $X\in R\otimes R$ in (\ref{oa}) and
\be
d_X = d_R^2 \cdot S_{\vac X}^2
\ee

\subsection{The case of $m=3$}

Quite similarly, for $m=3$ the pattern is

\begin{picture}(400,200)(-80,-120)
\put(0,0){\line(1,0){30}}
\put(0,0){\line(0,1){60}}
\put(0,60){\line(1,0){30}}
\put(30,0){\line(0,1){60}}
\put(2,2){\line(1,0){26}}
\put(2,2){\line(0,1){56}}
\put(2,58){\line(1,0){26}}
\put(28,2){\line(0,1){56}}
\put(-100,0){\line(1,0){30}}
\put(-100,0){\line(0,1){60}}
\put(-100,60){\line(1,0){30}}
\put(-70,0){\line(0,1){60}}
\put(-98,2){\line(1,0){26}}
\put(-98,2){\line(0,1){56}}
\put(-98,58){\line(1,0){26}}
\put(-72,2){\line(0,1){56}}
\put(50,0){\line(1,0){30}}
\put(50,0){\line(0,1){60}}
\put(50,60){\line(1,0){30}}
\put(80,0){\line(0,1){60}}
\put(52,2){\line(1,0){26}}
\put(52,2){\line(0,1){56}}
\put(52,58){\line(1,0){26}}
\put(78,2){\line(0,1){56}}
\put(100,0){\line(1,0){30}}
\put(100,0){\line(0,1){60}}
\put(100,60){\line(1,0){30}}
\put(130,0){\line(0,1){60}}
\put(102,2){\line(1,0){26}}
\put(102,2){\line(0,1){56}}
\put(102,58){\line(1,0){26}}
\put(128,2){\line(0,1){56}}
\put(200,0){\line(1,0){30}}
\put(200,0){\line(0,1){60}}
\put(200,60){\line(1,0){30}}
\put(230,0){\line(0,1){60}}
\put(202,2){\line(1,0){26}}
\put(202,2){\line(0,1){56}}
\put(202,58){\line(1,0){26}}
\put(228,2){\line(0,1){56}}
\put(260,20){\mbox{$\ldots$}}
\put(300,0){\line(1,0){30}}
\put(300,0){\line(0,1){60}}
\put(300,60){\line(1,0){30}}
\put(330,0){\line(0,1){60}}
\put(302,2){\line(1,0){26}}
\put(302,2){\line(0,1){56}}
\put(302,58){\line(1,0){26}}
\put(328,2){\line(0,1){56}}
\put(-50,-35){\line(1,0){30}}
\put(-50,-35){\line(0,-1){60}}
\put(-50,-95){\line(1,0){30}}
\put(-20,-35){\line(0,-1){60}}
\put(-48,-37){\line(1,0){26}}
\put(-48,-37){\line(0,-1){56}}
\put(-48,-93){\line(1,0){26}}
\put(-22,-37){\line(0,-1){56}}
\put(150,-35){\line(1,0){30}}
\put(150,-35){\line(0,-1){60}}
\put(150,-95){\line(1,0){30}}
\put(180,-35){\line(0,-1){60}}
\put(152,-37){\line(1,0){26}}
\put(152,-37){\line(0,-1){56}}
\put(152,-93){\line(1,0){26}}
\put(178,-37){\line(0,-1){56}}
\put(350,-35){\line(1,0){30}}
\put(350,-35){\line(0,-1){60}}
\put(350,-95){\line(1,0){30}}
\put(380,-35){\line(0,-1){60}}
\put(352,-37){\line(1,0){26}}
\put(352,-37){\line(0,-1){56}}
\put(352,-93){\line(1,0){26}}
\put(378,-37){\line(0,-1){56}}
\qbezier(-95,0)(-95,-15)(-95,-80)
\qbezier(-92,0)(-92,-18)(-92,-80)
%\put(-90,-20){\line(-1,0){45}}
%
\qbezier(-60,-21)(-89,-21)(-89,-60)
\qbezier(-89,-60)(-89,-100)(-60,-100)
\qbezier(-92,-80)(-92,-103)(-60,-103)
\qbezier(-95,-80)(-95,-106)(-60,-106)
\put(-60,-100){\line(1,0){450}}
\put(-60,-103){\line(1,0){450}}
\put(-60,-106){\line(1,0){450}}
\qbezier(420,-60)(420,-100)(390,-100)
\qbezier(423,-60)(423,-103)(390,-103)
\qbezier(426,-60)(426,-106)(390,-106)
\qbezier(325,0)(325,-15)(365,-15)\qbezier(395,-15)(426,-15)(426,-60)\put(365,-15){\line(1,0){30}}
\qbezier(375,-35)(375,-21)(395,-21)\qbezier(395,-21)(420,-21)(420,-60)
\qbezier(372,-35)(372,-18)(395,-18)\qbezier(395,-18)(423,-18)(423,-60)
\qbezier(-75,0)(-75,-15)(-35,-15)\qbezier(5,0)(5,-15)(-35,-15)
\qbezier(125,0)(125,-15)(165,-15)\qbezier(205,0)(205,-15)(165,-15)
\qbezier(225,0)(225,-15)(255,-15)\qbezier(305,0)(305,-15)(275,-15)
\qbezier(-45,-35)(-45,-21)(-60,-21)
\qbezier(-25,-35)(-25,-20)(15,-20)\qbezier(15,-20)(65,-20)(115,-20)\qbezier(155,-35)(155,-20)(115,-20)
\qbezier(175,-35)(175,-20)(215,-20)
\qbezier(-78,0)(-78,-18)(-60,-18)\qbezier(-42,-35)(-42,-18)(-60,-18)
\qbezier(8,0)(8,-18)(-10,-18)\qbezier(-28,-35)(-28,-18)(-10,-18)
\qbezier(122,0)(122,-18)(140,-18)\qbezier(158,-35)(158,-18)(140,-18)
\qbezier(208,0)(208,-18)(190,-18)\qbezier(172,-35)(172,-18)(190,-18)
\qbezier(322,0)(322,-18)(340,-18)\qbezier(358,-35)(358,-18)(340,-18)
\qbezier(25,0)(25,-15)(40,-15)\qbezier(55,0)(55,-15)(40,-15)
\qbezier(75,0)(75,-15)(90,-15)\qbezier(105,0)(105,-15)(90,-15)
\qbezier(22,0)(22,-17.5)(40,-17.5)\qbezier(58,0)(58,-17.5)(40,-17.5)
\qbezier(72,0)(72,-17.5)(90,-17.5)\qbezier(108,0)(108,-17.5)(90,-17.5)
\qbezier(222,0)(222,-17.5)(255,-17.5)\qbezier(308,0)(308,-17.5)(275,-17.5)
%\qbezier(255,-17.5)(255,-17.5)(255,-17.5)\qbezier(215,-20)(240,-20)(255,-20)
\qbezier(355,-35)(355,-20)(315,-20)\qbezier(275,-20)(295,-20)(315,-20)
%\qbezier(108,0)(108,-18)(65,-18)
%
%\qbezier(125,0)(125,-15)(165,-15)
%\qbezier(255,0)(255,-15)(215,-15)
%\qbezier(122,0)(122,-18)(165,-18)
%\qbezier(258,0)(258,-18)(215,-18)
%
%\qbezier(5,0)(5,-30)(60,-30)\qbezier(60,-30)(100,-30)(165,-30)
%\qbezier(275,0)(275,-30)(215,-30)
%\qbezier(8,0)(8,-27)(60,-27)\qbezier(60,-27)(100,-27)(165,-27)
%\qbezier(272,0)(272,-27)(215,-27)
%
%\put(5,0){\line(-1,-2){9.5}}
%\put(8,0){\line(-1,-2){10}}
%\put(25,0){\line(1,-2){9.5}}
%\put(22,0){\line(1,-2){10}}
%\put(5,60){\line(-1,2){9.5}}
%\put(8,60){\line(-1,2){10}}
%\put(25,60){\line(1,2){9.5}}
%\put(22,60){\line(1,2){10.0}}
%\put(63,-10){\mbox{$X$}}
%\put(160,-10){\mbox{$X$}}
%\put(220,-10){\mbox{$X$}}
%\put(0,-27){\mbox{$X$}}
%\put(-15,80){\mbox{$Y$}}
%\put(38,80){\mbox{$\bar Y$}}
\put(-95,28){\mbox{$B_{\vac X_1}^{(0)}$}}
\put(5,28){\mbox{$B_{\vac X_3}^{(1)}$}}
\put(55,28){\mbox{$B_{\vac X_3}^{(2)}$}}
\put(105,28){\mbox{$B_{\vac X_3}^{(3)}$}}
\put(205,28){\mbox{$B_{\vac X_5}^{(4)}$}}
\put(305,28){\mbox{$B_{\vac X_{s}}^{(g)}$}}
\put(-45,-62){\mbox{$B_{\vac X_2}^{(0')}$}}
\put(155,-62){\mbox{$B_{\vac X_4}^{(1')}$}}
\put(353,-62){\mbox{$B_{\vac X_{s'}}^{(g')}$}}
\put(60,-94){\mbox{$Q$}}
\put(-70,-40){\line(1,2){20}} \put(-60,5){\mbox{${\cal U}_{X_1X_2}$}}
\put(0,-40){\line(-1,2){20}} \put(5,-55){\mbox{${\cal U}_{X_2X_3}^\dagger$}}
\put(130,-40){\line(1,2){20}} \put(115,-55){\mbox{${\cal U}_{X_3X_4}$}}
\put(200,-40){\line(-1,2){20}} \put(200,-55){\mbox{${\cal U}_{X_4X_5}^\dagger$}}
\put(330,-40){\line(1,2){20}} \put(305,-55){\mbox{${\cal U}_{X_{s}X_{s'}}$}}
\put(400,-40){\line(-1,2){20}} \put(380,5){\mbox{${\cal U}_{X_{s'}X_1}^\dagger$}}
\end{picture}

\noindent
\paragraph{Pretzel fingers.} For the pretzel fingers
$B^{(p)}=P^{(n_p)}$ this figure
is equivalent to a knot lying on the surface of genus $g+g'$:

\begin{picture}(300,200)(-130,-113)
\put(-60,0){\line(1,0){20}}
\put(-60,0){\line(0,1){50}}
\put(-60,50){\line(1,0){20}}
\put(-40,0){\line(0,1){50}}
\put(-30,-30){\line(1,0){20}}
\put(-30,-30){\line(0,-1){50}}
\put(-30,-80){\line(1,0){20}}
\put(-10,-30){\line(0,-1){50}}
\put(0,0){\line(1,0){20}}
\put(0,0){\line(0,1){50}}
\put(0,50){\line(1,0){20}}
\put(20,0){\line(0,1){50}}
\put(30,0){\line(1,0){20}}
\put(30,0){\line(0,1){50}}
\put(30,50){\line(1,0){20}}
\put(50,0){\line(0,1){50}}
\put(60,0){\line(1,0){20}}
\put(60,0){\line(0,1){50}}
\put(60,50){\line(1,0){20}}
\put(80,0){\line(0,1){50}}
\put(90,-30){\line(1,0){20}}
\put(90,-30){\line(0,-1){50}}
\put(90,-80){\line(1,0){20}}
\put(110,-30){\line(0,-1){50}}
\put(120,0){\line(1,0){20}}
\put(120,0){\line(0,1){50}}
\put(120,50){\line(1,0){20}}
\put(140,0){\line(0,1){50}}
\put(165,10){\mbox{$\ldots$}}
\put(200,0){\line(1,0){20}}
\put(200,0){\line(0,1){50}}
\put(200,50){\line(1,0){20}}
\put(220,0){\line(0,1){50}}
\put(230,-30){\line(1,0){20}}
\put(230,-30){\line(0,-1){50}}
\put(230,-80){\line(1,0){20}}
\put(250,-30){\line(0,-1){50}}
\qbezier(-45,0)(-35,-15)(-25,-30)
\qbezier(5,0)(-5,-15)(-15,-30)
\qbezier(75,0)(85,-15)(95,-30)
\qbezier(125,0)(115,-15)(105,-30)
\qbezier(135,0)(143,-12)(143,-12)
\qbezier(197,-12)(205,0)(205,0)
\qbezier(215,0)(225,-15)(235,-30)
\qbezier(-45,50)(-45,60)(-20,60)\qbezier(5,50)(5,60)(-20,60)
\qbezier(15,50)(15,60)(25,60)\qbezier(35,50)(35,60)(25,60)
\qbezier(45,50)(45,60)(55,60)\qbezier(65,50)(65,60)(55,60)
\qbezier(15,0)(15,-10)(25,-10)\qbezier(35,0)(35,-10)(25,-10)
\qbezier(45,0)(45,-10)(55,-10)\qbezier(65,0)(65,-10)(55,-10)
\qbezier(75,50)(75,60)(100,60)
\qbezier(125,50)(125,60)(100,60)
\qbezier(135,50)(135,60)(160,60)
\qbezier(205,50)(205,60)(180,60)
\qbezier(-15,-80)(-15,-90)(10,-90)\qbezier(10,-90)(40,-90)(70,-90)\qbezier(95,-80)(95,-90)(70,-90)
\qbezier(105,-80)(105,-90)(130,-90)
\qbezier(235,-80)(235,-90)(210,-90)
\qbezier(210,-90)(200,-90)(180,-90)
\qbezier(130,-90)(150,-90)(160,-90)
\qbezier(-55,50)(-55,70)(-20,70)
\qbezier(215,50)(215,70)(180,70)
\qbezier(-20,70)(110,70)(180,70)
\qbezier(245,-80)(245,-100)(210,-100)
\qbezier(-25,-80)(-25,-100)(10,-100)
\qbezier(10,-100)(150,-100)(210,-100)
\put(-55,23){\mbox{$n_0$}}
\put(5,23){\mbox{$n_1$}}
\put(35,23){\mbox{$n_2$}}
\put(65,23){\mbox{$n_3$}}
\put(125,23){\mbox{$n_4$}}
\put(205,23){\mbox{$n_{g}$}}
\put(-25,-63){\mbox{$n_0'$}}
\put(95,-63){\mbox{$n_1'$}}
\put(234,-63){\mbox{$n_{g'}'$}}
\qbezier(-55,0)(-55,-15)(-30,-15)\qbezier(245,-30)(245,-15)(220,-15)\qbezier[100](-30,-15)(0,-15)(220,-15)
\end{picture}

\noindent
and in the case of the fundamental representation $R=[1]$ is described by the following expression:
$$
d_{[1]}H_{[1]} = d_{[3]} \cdot \prod_{i=0}^g P^{(n_i)}_{[2]}\prod_{j=0}^{g'} P^{(n_j')}_{[2]}
\ + \ d_{[111]}\cdot   \prod_{i=0}^g P^{(n_i)}_{[11]}\prod_{j=0}^{g'} P^{(n_j')}_{[11]} \ +
$$
\be
+\ d_{[21]}\cdot \Tr \Big( \Pi^{(n_0)}\ \Sigma\Pi^{(n'_0)}\Sigma\  \Pi^{(n_1)}\Pi^{(n_2)}\Pi^{(n_3)}
\ \Sigma \Pi^{(n'_1)}\Sigma\ \Pi^{(n_4)}\ \ldots \ \Pi^{(n_g)}\ \Sigma\Pi^{(n'_{g'})}\Sigma \Big)
\ee
where
\be
P^{(n)}_X =  \frac{(\bar S\bar T^{n}S)_{\vac X}}{S_{\vac X}}
\label{Pref}
\ee
and the second line contains $\mu_{[21]}\times\mu_{[21]}= 2\times 2$ matrices
\be
\Pi^{(n)} = \left(\begin{array}{cc} P^{(n)}_{[2]} & 0 \\ \\ 0 & P^{(n)}_{[11]}\end{array} \right)
\ \ \ \ \text{and} \ \ \ \
\Sigma={\cal U}_2 = \left(\begin{array}{cc} \frac{1}{[2]}& \frac{\sqrt{[3]}}{[2]} \\ \\
\frac{\sqrt{[3]}}{[2]} & -\frac{1}{[2]} \end{array}\right)
\ee
This formula is for the case of {\bf odd antiparallel} fingers, when all the three horizontal
strands are co-directed, and the mixing matrix ${\cal U}_2$ does
not depend on $N$.

\section{Examples. Three strands, $m=3$}

\subsection{Pretzel fingers}

Consider as a simple example the genus-five configuration
\be\label{dia}
H^{n_0,n_1,n_2,n_3,n_4,n_5} = "{\rm tr}"\ P_{n_0}P_{n_1}P_{n_2}UP_{n_3}UP_{n_4}UP_{n_5}U
\ee
made from six odd antiparallel pretzel fingers attached to a 3-strand braid in the following
sequence:  three consequent fingers to the first two stands, the forth one to the second two,
the fifth one to the first two and the sixth one to the second two:

\begin{picture}(300,100)(-70,-50)
\put(0,30){\vector(1,0){35}}
\put(0,0){\vector(1,0){35}}
\put(0,-30){\vector(1,0){35}}
\put(30,-30){\vector(1,0){125}}
\qbezier(35,30)(40,30)(40,25)
\qbezier(35,0)(40,0)(40,5)
\put(35,25){\line(1,0){20}}
\put(35,5){\line(0,1){20}}
\put(40,12){\mbox{$n_0$}}
\put(35,5){\line(1,0){20}}
\put(55,5){\line(0,1){20}}
\qbezier(50,25)(50,30)(55,30)
\qbezier(50,5)(50,0)(55,0)
\put(55,30){\vector(1,0){20}}
\put(55,0){\vector(1,0){20}}
\qbezier(75,30)(80,30)(80,25)
\qbezier(75,0)(80,0)(80,5)
\put(75,25){\line(1,0){20}}
\put(75,5){\line(0,1){20}}
\put(80,12){\mbox{$n_1$}}
\put(75,5){\line(1,0){20}}
\put(95,5){\line(0,1){20}}
\qbezier(90,25)(90,30)(95,30)
\qbezier(90,5)(90,0)(95,0)
\put(95,30){\vector(1,0){20}}
\put(95,0){\vector(1,0){20}}
\qbezier(115,30)(120,30)(120,25)
\qbezier(115,0)(120,0)(120,5)
\put(115,25){\line(1,0){20}}
\put(115,5){\line(0,1){20}}
\put(120,12){\mbox{$n_2$}}
\put(115,5){\line(1,0){20}}
\put(135,5){\line(0,1){20}}
\qbezier(130,25)(130,30)(135,30)
\qbezier(130,5)(130,0)(135,0)
\put(135,30){\vector(1,0){20}}
\put(135,0){\vector(1,0){20}}
\qbezier(155,-30)(160,-30)(160,-25)
\qbezier(155,0)(160,0)(160,-5)
\put(155,-25){\line(1,0){20}}
\put(155,-5){\line(0,-1){20}}
\put(160,-17){\mbox{$n_3$}}
\put(155,-5){\line(1,0){20}}
\put(175,-5){\line(0,-1){20}}
\qbezier(170,-25)(170,-30)(175,-30)
\qbezier(170,-5)(170,0)(175,0)
\put(155,30){\line(1,0){40}}
\put(175,-30){\vector(1,0){60}}
\put(175,0){\vector(1,0){20}}
\qbezier(195,30)(200,30)(200,25)
\qbezier(195,0)(200,0)(200,5)
\put(195,25){\line(1,0){20}}
\put(195,5){\line(0,1){20}}
\put(200,12){\mbox{$n_4$}}
\put(195,5){\line(1,0){20}}
\put(215,5){\line(0,1){20}}
\qbezier(210,25)(210,30)(215,30)
\qbezier(210,5)(210,0)(215,0)
\put(215,30){\vector(1,0){70}}
\put(215,0){\vector(1,0){20}}
\qbezier(235,-30)(240,-30)(240,-25)
\qbezier(235,0)(240,0)(240,-5)
\put(235,-25){\line(1,0){20}}
\put(235,-5){\line(0,-1){20}}
\put(240,-17){\mbox{$n_5$}}
\put(235,-5){\line(1,0){20}}
\put(255,-5){\line(0,-1){20}}
\qbezier(250,-25)(250,-30)(255,-30)
\qbezier(250,-5)(250,0)(255,0)
\put(255,-30){\vector(1,0){30}}
\put(255,0){\vector(1,0){30}}
%\put(235,30){\vector(1,0){20}}
%
\end{picture}

Already this family contains all but one knots with up to seven intersections
(only one example per knot is given):
\be
\begin{array}{c|c}
{\rm knot} &  n_0,n_1,n_2,n_3,n_4,n_5  \\
 \hline
3_1 &  1, 1, 1, 1, 1, -1   \\
4_1 &  -3, 1, 1, -1, 5, 1 \\
5_1& 1, 1, 1, 1, 1, 1  \\
5_2& 1, 1, 1, 1, -1, 1   \\
6_1 & 1, 3, -3, -1, 3, 1 \\
6_2& 1, 1, 1, -1, 1, -1  \\
6_3 & 1, 1, 1, -1, -3, -1 \\
7_2 & 1, 3, -1, 1, 3, 1 \\
7_3 & 3, 1, 1, 1, 1, 1 \\
7_4 & 1, 3, 3, -1, -3, 1 \\
7_5 & 1, 1, 1, 1, 3, 1 \\
7_6 & 1, 1, 1, 1, -3, 1 \\
7_7 & -1, 3, 3, -3, -1, 1\\
\hline
\end{array}
\ee
\be
\begin{array}{c|c}
\hline
8_1& -1, 3, -1, -1, 1, -3\\
8_3 & -5, 1, 3, -1, 3, 1 \\
8_4 & -1, -1, -1, 3, -1, 1\\
8_6 & -1, -1, -1, 5, -3, -3 \\
%8_8 ????????&  -1, -1, -1, 1, 5, 1 \\
8_{11} & -1, 3, -1, -1, -1, -3\\
8_{12} & -1, 3, -1, -1, 3, -1\\
8_{13} &  -3, -1, -1, 5, 3, -3 \\
8_{14} & -1, 3, -1, 1, -3, -3\\
\hline
9_2 & 1, 3, -1, 5, -1, 3 \\
9_4 & 5, 1, 1, 1, 1, 1\\
9_5 &  -3, -1, 1, -1, -3, -3 \\
9_7& 1, 1, 1, 1, 5, 1 \\
9_8 & -1, -1, -1, 1, 5, -3 \\
9_{10} & -3,-3,-1,-3,-1,1\\
9_{12} &  -1, -1, 5, 5, 3, -3 \\
9_{13} & 1, 1, 1, 1, 3, 3 \\
9_{14} &   -1, 3, -1, 3, -1, 1 \\
9_{15}& -1, 3, -1, 5, 5, -3 \\
9_{18}& 1, 3, 1, 1, 3, 1 \\
9_{19} &  -1, 3, -1, 3, -3, -1 \\
9_{21}& 1, 3, 1, 1, -3, 1\\
9_{35} & -3, -1, -3, -1, 1, -1\\
9_{37}&-1, 3, -3, 1, -1, 1\\
9_{46} & -3, -1, 3, -1, 1, -1\\
9_{48} & -1, 3, 3, -3, -1, 5 \\
\hline
\end{array}
\ee

\subsection{Non-pretzel finger}

One can further substitute pretzel fingers in (\ref{dia}) by non-pretzel ones.
The simplest non-pretzel finger is the parallel-antiparallel braid of \cite{evo},

\begin{picture}(300,100)(-70,-40)
\put(20,30){\vector(1,0){15}}
\put(20,0){\vector(1,0){15}}
\put(55,30){\vector(1,0){15}}
\put(55,0){\vector(1,0){15}}
\qbezier(35,30)(40,30)(40,25)
\qbezier(35,0)(40,0)(40,5)
\put(35,25){\line(1,0){20}}
\put(35,5){\line(0,1){20}}
\put(43,12){\mbox{$n$}}
\put(35,5){\line(1,0){20}}
\put(55,5){\line(0,1){20}}
\qbezier(50,25)(50,30)(55,30)
\qbezier(50,5)(50,0)(55,0)
\put(85,15){\mbox{$\longrightarrow$}}
\put(110,40){\vector(1,0){25}}
\put(110,-20){\vector(1,0){25}}
\put(155,40){\vector(1,0){25}}
\put(155,-20){\vector(1,0){25}}
\qbezier(135,30)(125,30)(125,20)
\qbezier(135,10)(125,10)(125,20)
\qbezier(135,10)(140,10)(140,5)
\qbezier(135,-20)(140,-20)(140,-15)
\put(135,5){\line(1,0){20}}
\put(135,-15){\line(0,1){20}}
\put(143,-8){\mbox{$n$}}
\put(135,-15){\line(1,0){20}}
\put(155,-15){\line(0,1){20}}
\qbezier(150,5)(150,10)(155,10)
\qbezier(150,-15)(150,-20)(155,-20)
\qbezier(165,20)(165,10)(155,10)
\qbezier(165,20)(165,30)(155,30)
\put(135,25){\line(1,0){20}}
\put(135,25){\line(0,1){20}}
\put(140,30){\mbox{$2m$}}
\put(135,45){\line(1,0){20}}
\put(155,25){\line(0,1){20}}
\end{picture}

\noindent
what means that $P^{(n)}_X =  \frac{(\bar S\bar T^{n}S)_{\vac X}}{S_{\vac X}}$
from (\ref{Pref}) with odd $n$ is changed for
\be
{\rm either} \ \ \ \ \ \
K^{(m,n)}_X = \frac{( S T^{2m} S^\dagger \bar T^{n}S)_{\vac X}}{S_{\vac X}}
 \ \ \ \ \ \ {\rm or} \ \ \ \ \ \
 \bar K^{(m,n)}_X = \frac{(\bar S\bar T^{2m}\bar S \bar T^{n}S)_{\vac X}}{S_{\vac X}}
\ee
with $m\neq 0$ and $n$ either even or odd respectively.

For an example involving these fingers see the Appendix at the end of this paper.
Note that at the level of symmetric HOMFLY polynomials one can safely permute horizontal and
vertical braids in the above picture: this is a mutation transform affecting only
$H_{[21]}$ and other non-trivially colored polynomials.

\section{Generalized composites}

The set of knots, which can be handled by the above method,
is further enlarged by inclusion of generalized composites.

\subsection{Ordinary composite knots}

The ordinary composite knot looks like

\begin{picture}(300,70)(-150,-30)
%\put(0,0){\line(1,0){40}}
\put(10,0){\circle{20}}
%\put(80,0){\line(1,0){40}}
\put(40,0){\circle{20}}
%\put(160,0){\line(1,0){40}}
\qbezier(10,10)(10,25)(25,25)
\qbezier(40,10)(40,25)(25,25)
\qbezier(10,-10)(10,-25)(25,-25)
\qbezier(40,-10)(40,-25)(25,-25)
\put(2,-4){\mbox{$B^{(1)}$}}
\put(32,-4){\mbox{$B^{(2)}$}}
\put(60,0){\mbox{$\longleftarrow$}}
\put(100,0){\circle{20}}
\qbezier(100,10)(100,25)(115,25)
\qbezier(130,10)(130,25)(115,25)
\qbezier(100,-10)(100,-25)(115,-25)
\qbezier(130,-10)(130,-25)(115,-25)
\put(130,-10){\line(0,1){20}}
\put(92,-4){\mbox{$B^{(1)}$}}
\put(190,0){\circle{20}}
\qbezier(190,10)(190,25)(175,25)
\qbezier(160,10)(160,25)(175,25)
\qbezier(190,-10)(190,-25)(175,-25)
\qbezier(160,-10)(160,-25)(175,-25)
\put(160,-10){\line(0,1){20}}
\put(182,-4){\mbox{$B^{(2)}$}}
\end{picture}

\noindent
and the main fact is that the reduced HOMFLY for it is a product:
\be
H_R^{\text{comp}}  = H^{(1)}_RH^{(2)}_R
\label{ordcomp}
\ee
This equality comes from the simple fact: the "open" graph
is a unit matrix in representation $R$ times a factor $B$,
and the unreduced HOMFLY polynomial is its graded trace $d_RH_R = B\cdot{\rm tr}_R I = d_RB$.
Matrix for a composite is a product of matrices, thus  $B=B^{(1)}B^{(2)}$
and (\ref{ordcomp}) is a corollary.

\subsection{2-composites}

Similarly, for the graph

\begin{picture}(300,80)(-150,-30)
%\put(0,0){\line(1,0){40}}
\put(10,0){\circle{20}}
%\put(80,0){\line(1,0){40}}
\put(40,0){\circle{20}}
%\put(160,0){\line(1,0){40}}
\qbezier(9,10)(9,26)(25,26)
\qbezier(41,10)(41,26)(25,26)
\qbezier(9,-10)(9,-26)(25,-26)
\qbezier(41,-10)(41,-26)(25,-26)
\qbezier(11,10)(11,24)(25,24)
\qbezier(39,10)(40,24)(25,24)
\qbezier(11,-10)(11,-24)(25,-24)
\qbezier(39,-10)(39,-24)(25,-24)
\put(2,-4){\mbox{$B^{(1)}$}}
\put(32,-4){\mbox{$B^{(2)}$}}
\put(60,0){\mbox{$\longleftarrow$}}
\put(100,0){\circle{20}}
\qbezier(99,10)(99,26)(115,26)
\qbezier(131,10)(131,26)(115,26)
\qbezier(99,-10)(99,-26)(115,-26)
\qbezier(131,-10)(131,-26)(115,-26)
\put(125,-10){\line(0,1){20}}
\qbezier(101,10)(101,24)(115,24)
\qbezier(129,10)(129,24)(115,24)
\qbezier(101,-10)(101,-24)(115,-24)
\qbezier(129,-10)(129,-24)(115,-24)
\put(135,-10){\line(0,1){20}}
\put(125,10){\line(1,0){10}}
\put(125,-10){\line(1,0){10}}
\put(126,-2){\mbox{$k_1$}}
\put(92,-4){\mbox{$B^{(1)}$}}
\put(190,0){\circle{20}}
\qbezier(191,10)(191,26)(175,26)
\qbezier(159,10)(159,26)(175,26)
\qbezier(191,-10)(191,-26)(175,-26)
\qbezier(159,-10)(159,-26)(175,-26)
\put(155,-10){\line(0,1){20}}
\qbezier(189,10)(189,24)(175,24)
\qbezier(161,10)(160,24)(175,24)
\qbezier(189,-10)(189,-24)(175,-24)
\qbezier(161,-10)(160,-24)(175,-24)
\put(165,-10){\line(0,1){20}}
\put(155,10){\line(1,0){10}}
\put(155,-10){\line(1,0){10}}
\put(156,-2){\mbox{$k_2$}}
\put(182,-4){\mbox{$B^{(2)}$}}
\put(20,33){\mbox{$H_R$}}
\put(105,33){\mbox{$H_R^{k_1,(1)}$}}
\put(165,33){\mbox{$H_R^{k_2,(2)}$}}
\end{picture}

\noindent
with two strands between the blobs, the same factorization holds, $B=B^{(1)}B^{(2)}$, but now
$B$ is a diagonal matrix, with entries $B_Q$, $Q\in R\otimes R$ and
\be
d_RH_R^{2\text{-comp}} = \sum_{Q\in R\otimes R} d_QB_Q = \sum_{Q\in R\otimes R} d_QB_Q^{(1)}B_Q^{(2)}
\ee
The HOMFLY constituents
\be
d_RH_R^{(i)}=\sum_{Q\in R\otimes R} d_QB_Q^{(i)}
\ee
define only particular linear combinations of $B_Q^{(i)}$, but one can insert additional
${\cal R}$-matrices (a two-strand braid of length $k_i$) to extract an additional information.
For example, in the case of the fundamental representation $R=[1]$, there are just two $Q=[2],[11]$
and just two values choices $k_i=0,1$ are sufficient to extract $B_{[2]}^{(i)}$ and $B_{[11]}^{(i)}$:
\be
d_{[2]}B_{[2]}^{(i)} +d_{[11]}B_{[11]}^{(i)} = d_{[1]}H_{[1]}^{(i)}, \nn \\
d_{[2]}\lambda_{[2]}B_{[2]}^{(i)} +d_{[11]}\lambda_{[11]}B_{[11]}^{(i)} = d_{[1]}H_{[1]}^{1,(i)}
\ee
where $\lambda_Q$ are eigenvalues of the ${\cal R}$-matrix
and $H^{1,(i)}$ is the HOMFLY polynomial of the closure of the blob $B^{(i)}$ with one additional intersection:
$k_i=1$.
Substituting $d_{[1]}=\frac{\{A\}}{\{q\}}$,  $d_{[2]} = \frac{\{A\}\{Aq\}}{\{q\}\{q^2\}}$,
and $d_{[11]}=\frac{\{A\}\{A/q\}}{\{q\}\{q^2\}}$ (we use the notation $\{x\}\equiv x-1/x$) and $\lambda_{[2]}=\frac{q}{A}$,
$\lambda_{[11]}=-\frac{1}{qA}$, one gets for the reduced HOMFLY polynomial
\be
H_{[1]}^{2\text{-comp}} =
\frac{d_{[2]}B_{[2]}^{(1)}B_{[2]}^{(2)}+ d_{[11]}B_{[11]}^{(1)}B_{[11]}^{(2)}}{d_{[1]}}
= \frac{\Big(A(q^2-1+q^{-2})-A^{-1}\Big)\{q\}}{\{Aq\}\{A/q\}}\cdot H_{[1]}^{(1)}H_{[1]}^{(2)} + \nn \\
+\frac{A^2\{A\}\{q\}}{\{Aq\}\{A/q\}}\cdot H_{[1]}^{1,(1)}H_{[1]}^{1,(2)}
- \frac{A^2\{q\}^2}{\{Aq\}\{A/q\}}\Big(H_{[1]}^{(1)}H_{[1]}^{1,(2)}+H_{[1]}^{1,(1)}H_{[1]}^{(2)}\Big)
\label{2comp}
\ee
For $R\neq [1]$ the final formula involves more braids with different lengths $k_i$.

\subsection{$m$-composites}

Similarly one can handle a composition $H_R^{m\text{-comp}}$ of two blobs
connected by arbitrary number $m$ of strands.
For additional $m$-strand braids one can (but is not obliged to) take just torus ones of lengths $k_i$.

\subsection{Examples of (\ref{2comp})}

First, consider the case when both $B^{(i)}$ are composites themselves:

\begin{picture}(300,80)(-150,-40)
\put(-10,0){\circle{20}}
%\put(80,0){\line(1,0){40}}
\put(60,0){\circle{20}}
%\put(160,0){\line(1,0){40}}
\qbezier(-10,10)(-10,25)(5,25)
\qbezier(60,10)(60,25)(45,25)
\put(5,25){\line(1,0){40}}
\put(5,-25){\line(1,0){40}}
\qbezier(-10,-10)(-10,-25)(5,-25)
\qbezier(60,-10)(60,-25)(45,-25)
\put(-18,-4){\mbox{$B^{(1)}$}}
\put(52,-4){\mbox{$B^{(2)}$}}
\put(25,0){\circle{40}}
\put(97,0){\mbox{or}}
\put(150,0){\circle{20}}
%\put(80,0){\line(1,0){40}}
\put(195,0){\circle{20}}
%\put(160,0){\line(1,0){40}}
\qbezier(150,10)(150,35)(175,35)
\qbezier(210,10)(210,35)(185,35)
\put(175,35){\line(1,0){10}}
\put(175,-35){\line(1,0){10}}
\qbezier(150,-10)(150,-35)(175,-35)
\qbezier(210,-10)(210,-35)(185,-35)
\qbezier(165,10)(165,25)(180,25)
\qbezier(195,10)(195,25)(180,25)
\put(165,-10){\line(0,1){20}}
\put(210,-10){\line(0,1){20}}
\qbezier(165,-10)(165,-25)(180,-25)
\qbezier(195,-10)(195,-25)(180,-25)
\put(142,-4){\mbox{$B^{(1)}$}}
\put(187,-4){\mbox{$B^{(2)}$}}
\end{picture}

\noindent
In this case $d_RH^{(i)}_R = d_R^2B^{(i)}_R$ and $d_RH^{1,(i)}_R = d_R B^{(i)}_R$,
and substituting this into (\ref{2comp}) one gets
$H^{2\text{-comp}}_{[1]} = d_{[1]}B^{(1)}B^{(2)}=d_{[1]}^{-1}H^{(1)}_{[1]}H^{(2)}_{[2]}$,
what is the right answer (note that the HOMFLY polynomials in this case are associated with links and
all are defined to contain an extra unknot factor $d_{[1]}$).

\bigskip

More interesting is the case, when the blobs are obtained by closing one line in
3-strand braids:

\begin{picture}(300,100)(-150,-50)
\put(0,0){\circle{20}}
\put(-8,-4){\mbox{$B^{(i)}$}}
\put(-10,1){\line(-1,0){20}}
\put(-10,-1){\line(-1,0){20}}
\put(10,1){\line(1,0){20}}
\put(10,-1){\line(1,0){20}}
\put(55,-1){\mbox{$=$}}
\put(90,1){\line(1,0){20}}
\put(90,-1){\line(1,0){20}}
\put(110,-10){\line(0,1){25}}
\put(110,-10){\line(1,0){35}}
\put(110,15){\line(1,0){35}}
\put(145,15){\line(0,-1){25}}
\put(145,1){\line(1,0){20}}
\put(145,-1){\line(1,0){20}}
\qbezier(110,5)(105,5)(105,10)
\qbezier(110,25)(105,25)(105,20)
\qbezier(145,5)(150,5)(150,10)
\qbezier(145,25)(150,25)(150,20)
\put(110,25){\line(1,0){35}}
\put(105,10){\line(0,1){10}}
\put(150,10){\line(0,1){10}}
\put(112,-1){\mbox{$\text{braid}^{(i)}$}}
\end{picture}

\vspace{-0.5cm}

\noindent
An example of this type is provided by the mutant $11a19$ given by the following closed braid:

%\bigskip
\vspace{-0.5cm}
\hspace{1cm}
\begin{figure}[h!]
\centering\leavevmode
\includegraphics[width=8 cm]{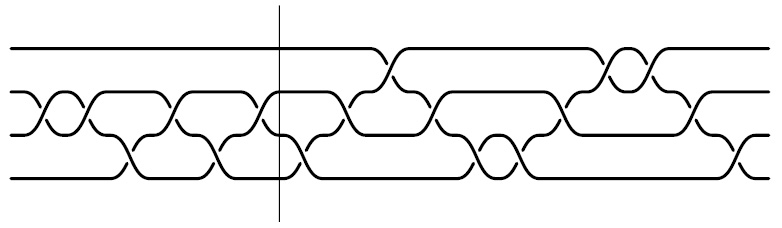}
\label{example}
\end{figure}
\vspace{-0.5cm}
%\bigskip

\noindent
which can be redrawn as

%\bigskip

\begin{figure}[h!]
\centering\leavevmode
\includegraphics[width=8 cm]{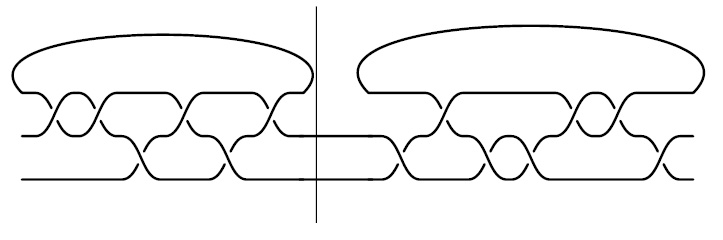}
\label{example}
\end{figure}
\vspace{-0.7cm}

%\bigskip

\noindent
Thus, this braid can be presented with the two constituent sub-braids (with the topmost strand closed):

\bigskip

\noindent
braid$^{(1)}$

%\bigskip
\vspace{-0.6cm}
\hspace{4cm}
\begin{figure}[h!]
\centering\leavevmode
\includegraphics[width=4 cm]{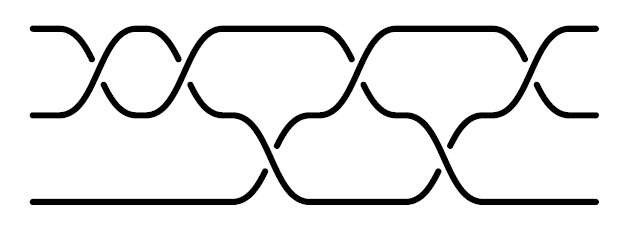}
\label{example}
\end{figure}

\vspace{-0.6cm}
%\bigskip

\noindent
braid$^{(2)}$

%\bigskip
\vspace{-0.5cm}
\hspace{4cm}
\begin{figure}[h!]
\centering\leavevmode
\includegraphics[width=4 cm]{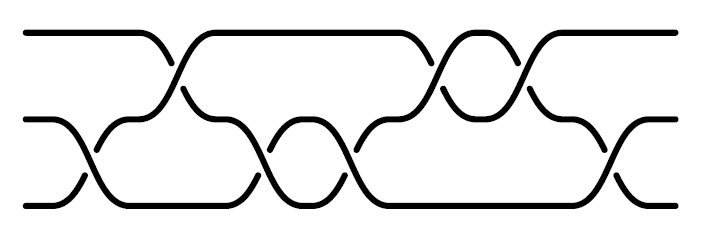}
\label{example}
\end{figure}

%\vspace{-0.5cm}

%\bigskip

\noindent
Then, these constituents are equal to
\be
B^{(1)}_{[2]}={2A^2q^6-q^8-2A^2q^4+2q^6+2A^2q^2-2q^4-A^2+q^2\over q^5A^3}\nn\\
B^{(1)}_{[11]}={A^2q^8-2A^2q^6+2A^2q^4-q^6-2A^2q^2+2q^4-2q^2+1\over A^3q^3}\nn\\
B^{(2)}_{[2]}={2A^2q^4-q^6-A^2q^2+q^4+A^2-q^2\over A^2q^4}\nn\\
B^{(2)}_{[11]}={A^2q^6-A^2q^4+2A^2q^2-q^4+q^2-1\over q^2A^2}
\ee
and eq.(\ref{2comp}) gives
\be
H_{[1]}^{11a19} ={1\over A^6q^8}\left(
A^6q^{14}-A^4q^{16}-3A^6q^{12}+3A^4q^{14}+7A^6q^{10}-9A^4q^{12}+2A^2q^{14}-7A^6q^8+13A^4q^{10}-\right.\nn\\
\phantom{{1\over A^6q^8}}-5A^2q^{12}+7A^6q^6-17A^4q^8+
11A^2q^{10}-q^{12}-3A^6q^4+13A^4q^6-12A^2q^8+2q^{10}+\nn\\\left.+A^6q^2-9A^4q^4+11A^2q^6-3q^8+3A^4q^2-5A^2q^4+2q^6-A^4+2A^2q^2-q^4\right)
\ee

\section{Conclusion}

In this paper, we further extended the method of \cite{MMMRS} to a wider class of knots by attaching "fingers" to pairs of
adjacent strands in closed braids and by considering "m-composites". We illustrated the story by examples of the
HOMFLY polynomials in the fundamental representation, generalization to (anti)symmetric representations is straightforward, extension to non-symmetric representations with additional degeneracies will be described elsewhere. Another subject to be considered separately is inclusion of "loops", marked as "forbidden" in (\ref{fig}).

\section*{Acknowledgements}

We are indebted to Yakov Kononov for help with the Appendix.
This work was performed at the
Institute for Information Transmission Problems with the financial support of the Russian Science
Foundation (Grant No.14-50-00150).

\newpage

\section{Appendix}

We present here a family, which includes almost all the up-to-10-crossings knots.
It is extremely convenient to have a whole family for testing various hypothesis, which are supposed to be
true universally, i.e. for all knots (like those in \cite{KoMo}): this allows one to generate the concrete colored HOMFLY polynomial
merely by choosing the proper integers $\{n_i\}$. Because of it, this is often much more convenient than using the sophisticated table in \cite{MMMRS}. In fact, there are a lot of such families, and this is no way distinguished among them
(and is even hardly the smallest of this kind). The family is parameterized by seven integers $\{n_i\}$ and looks like

\begin{picture}(300,100)(-20,-50)
\put(40,30){\vector(1,0){35}}
\put(40,0){\vector(1,0){35}}
\put(40,-30){\vector(1,0){35}}
\put(75,-30){\vector(1,0){80}}
\qbezier(80,25)(80,35)(85,35)\qbezier(85,35)(90,35)(90,25)
\put(86,35){\vector(1,0){2}}
\qbezier(75,0)(80,0)(80,5)
\put(75,25){\line(1,0){20}}
\put(75,5){\line(0,1){20}}
\put(80,12){\mbox{$n_1$}}
\put(75,5){\line(1,0){20}}
\put(95,5){\line(0,1){20}}
%\qbezier(90,25)(90,30)(95,30)
\qbezier(90,5)(90,0)(95,0)
\put(75,30){\vector(1,0){40}}
\put(95,0){\vector(1,0){20}}
\qbezier(115,30)(120,30)(120,25)
\qbezier(115,0)(120,0)(120,5)
\put(115,25){\line(1,0){20}}
\put(115,5){\line(0,1){20}}
\put(120,12){\mbox{$n_2$}}
\put(115,5){\line(1,0){20}}
\put(135,5){\line(0,1){20}}
\qbezier(130,25)(130,30)(135,30)
\qbezier(130,5)(130,0)(135,0)
\put(135,30){\vector(1,0){20}}
\put(135,0){\vector(1,0){20}}
\qbezier(155,-30)(160,-30)(160,-25)
\qbezier(155,0)(160,0)(160,-5)
\put(155,-25){\line(1,0){20}}
\put(155,-5){\line(0,-1){20}}
\put(160,-17){\mbox{$n_3$}}
\put(155,-5){\line(1,0){20}}
\put(175,-5){\line(0,-1){20}}
\qbezier(170,-25)(170,-30)(175,-30)
\qbezier(170,-5)(170,0)(175,0)
\put(155,30){\line(1,0){40}}
\put(175,-30){\vector(1,0){60}}
\put(175,0){\vector(1,0){20}}
\qbezier(195,30)(200,30)(200,25)
\qbezier(195,0)(200,0)(200,5)
\put(195,25){\line(1,0){20}}
\put(195,5){\line(0,1){20}}
\put(200,12){\mbox{$n_4$}}
\put(195,5){\line(1,0){20}}
\put(215,5){\line(0,1){20}}
\qbezier(210,25)(210,30)(215,30)
\qbezier(210,5)(210,0)(215,0)
\put(215,30){\vector(1,0){150}}
\put(215,0){\vector(1,0){20}}
\qbezier(235,-30)(240,-30)(240,-25)
\qbezier(235,0)(240,0)(240,-5)
\put(235,-25){\line(1,0){20}}
\put(235,-5){\line(0,-1){20}}
\put(240,-17){\mbox{$n_5$}}
\put(235,-5){\line(1,0){20}}
\put(255,-5){\line(0,-1){20}}
\qbezier(250,-25)(250,-30)(255,-30)
\qbezier(250,-5)(250,0)(255,0)
\put(255,-30){\vector(1,0){60}}
\put(255,0){\vector(1,0){20}}
%\put(235,30){\vector(1,0){20}}
%

\qbezier(280,25)(280,35)(285,35)\qbezier(285,35)(290,35)(290,25)
\put(286,35){\vector(1,0){2}}
\qbezier(275,0)(280,0)(280,5)
\put(275,25){\line(1,0){20}}
\put(275,5){\line(0,1){20}}
\put(280,12){\mbox{$n_6$}}
\put(275,5){\line(1,0){20}}
\put(295,5){\line(0,1){20}}
%\qbezier(90,25)(90,30)(95,30)
\qbezier(290,5)(290,0)(295,0)
%
%\put(275,30){\vector(1,0){80}}
\put(295,0){\vector(1,0){20}}
\qbezier(320,-25)(320,-35)(325,-35)\qbezier(325,-35)(330,-35)(330,-25)
\put(324,-35){\vector(-1,0){2}}
%\qbezier(315,-30)(320,-30)(320,-25)
\qbezier(315,0)(320,0)(320,-5)
\put(315,-25){\line(1,0){20}}
\put(315,-5){\line(0,-1){20}}
\put(320,-17){\mbox{$n_7$}}
\put(315,-5){\line(1,0){20}}
\put(335,-5){\line(0,-1){20}}
%\qbezier(330,-25)(330,-30)(335,-30)
\qbezier(330,-5)(330,0)(335,0)
%
%\put(315,30){\line(1,0){40}}
\put(315,-30){\vector(1,0){50}}
\put(335,0){\vector(1,0){30}}
\put(273,30){\vector(1,0){2}}
\end{picture}

\noindent
Here $n_1$ and $n_6$ are even, the other five parameters $n_{2,3,4,5}$ and $n_7$ are odd.
We also distinguish between the two ${\cal R}$-matrix orientations
in the small loop below the last box with $n_7$:
\be
P^{(n_{2,3,4,5})}_X =  \frac{(\bar S\bar T^{n_{2,3,4,5}}S)_{_{\vac,X}}}{S_{_{\vac, X}}}\nn\\
K^{(n_{1,6})}_X = \frac{( S T^2 S^\dagger \bar T^{n_{1,6}}S)_{_{\vac, X}}}{S_{_{\vac, X}}}\nn\\
\bar K^{(n_7|\pm)}_X = \frac{(\bar S\bar T^{\pm 2}\bar S \bar T^{n_7}S)_{_{\vac, X}}}{S_{_{\vac, X}}}
\label{7parfam}
\ee
This picture of the knot is rather symbolic, since one has also to mark the way how the small loops nearby
the boxes $n_1$, $n_6$ and $n_7$ cross the strands. It can be read off from the formula that is really used for the calculation:
in the case of the fundamental representation $R=[1]$:
$$
d_{[1]}H_{[1]}^{(n_1,\ldots,n_7|\pm)} \ \ = \ \ d_{[3]} \cdot
K^{(n_1)}_{[2]}\cdot \left(
\prod_{i=2}^{5}   P^{(n_i)}_{[2]}\right) K^{(n_6)}_{[2]}\bar K^{(n_7|\pm)}_{[2]}
\ \ + \ \ d_{[111]}\cdot    K^{(n_1)}_{[11]}\cdot \left(
\prod_{i=2}^{5}   P^{(n_i)}_{[11]}\right) K^{(n_6)}_{[11]}\bar K^{(n_7|\pm)}_{[11]} \ \ +
$$
\vspace{-0.3cm}
\begin{multline}
+\ \ d_{[21]}\cdot  \Tr_{2\times 2} \left\{
\left(\begin{array}{cc} K^{(n_1)}_{[2]} & 0 \\ \\  0 & K^{(n_1)}_{[11]} \end{array}\right)
\left(\begin{array}{cc} P^{(n_2)}_{[2]} & 0 \\ \\  0 & P^{(n_2)}_{[11]} \end{array}\right)
\left(\begin{array}{cc} \frac{1}{[2]}& \frac{\sqrt{[3]}}{[2]} \\ \\
\frac{\sqrt{[3]}}{[2]} & -\frac{1}{[2]} \end{array}\right)
\left(\begin{array}{cc} P^{(n_3)}_{[2]} & 0 \\ \\  0 & P^{(n_3)}_{[11]} \end{array}\right)
\left(\begin{array}{cc} \frac{1}{[2]}& \frac{\sqrt{[3]}}{[2]} \\ \\
\frac{\sqrt{[3]}}{[2]} & -\frac{1}{[2]} \end{array}\right)  \cdot
\right.
%$$
%$$
\\
 \cdot\left(\begin{array}{cc} P^{(n_4)}_{[2]} & 0 \\ \\  0 & P^{(n_4)}_{[11]} \end{array}\right)
\left(\begin{array}{cc} \frac{1}{[2]}& \frac{\sqrt{[3]}}{[2]} \\ \\
\frac{\sqrt{[3]}}{[2]} & -\frac{1}{[2]} \end{array}\right)
\left(\begin{array}{cc} P^{(n_5)}_{[2]} & 0 \\ \\  0 & P^{(n_5)}_{[11]} \end{array}\right)
\left(\begin{array}{cc} \frac{1}{[2]}& \frac{\sqrt{[3]}}{[2]} \\ \\
\frac{\sqrt{[3]}}{[2]} & -\frac{1}{[2]} \end{array}\right)
\left(\begin{array}{cc} K^{(n_6)}_{[2]} & 0 \\ \\  0 & K^{(n_6)}_{[11]} \end{array}\right)
\cdot
%$$
\\
\cdot
\left.
\left(\begin{array}{cc} \frac{1}{[2]}& \frac{\sqrt{[3]}}{[2]} \\ \\
\frac{\sqrt{[3]}}{[2]} & -\frac{1}{[2]} \end{array}\right)
\left(\begin{array}{cc} \bar K^{(n_7|\pm)}_{[2]} & 0 \\ \\  0 & \bar K^{(n_7|\pm)}_{[11]} \end{array}\right)
\left(\begin{array}{cc} \frac{1}{[2]}& \frac{\sqrt{[3]}}{[2]} \\ \\
\frac{\sqrt{[3]}}{[2]} & -\frac{1}{[2]} \end{array}\right)
 \right\} \ \ \ \ \ \ \ \
\end{multline}

\noindent
The possible representatives of knots are:
$$
\begin{array}{c|c|c}
{\rm knot} &   n_1,n_2,n_3,n_4, n_5,n_6,n_7 & \pm \\
 \hline
3_1 &  0, -1, 1, -1, -1, 0, 1 & +  \\
\hline
4_1 &  0, -1, 1, 1, 3, 0, -1 & + \\
\hline
5_1 & -4, -1, 1, 1, 1, -2, -1&  +  \\
5_2& 0, -1, -1, -1, 1, 0, 1 & + \\
\hline
6_1& 0, -1, 3, 1, 1, 0, -1 & + \\
6_2& 0, -1, -1, -1, 3, 2, 1 & + \\
6_3& 0, -1, 1, 1, 1, 0, 1 & + \\
\hline
7_1 & 0, -1, 1, 1, -1, 0, -3 & - \\
7_2 & 0, -1, -1, -1, 1, -2, 1 & +\\
7_3 &  0, 1, 1, 1, -3, 0, -3  & +\\
7_4 &  0, 1, -3, -1, -1, 0, 1 & +   \\
7_5 & 0, -1, -1, -1, -1, 0, 1 & + \\
7_6 & 0, 1, -1, 1, -1, 0, -1 & + \\
7_7& 0, -1, 1, 1, 1, 2, -1 & +  \\
\hline
8_1 & 0, 1, -1, -1, 3, 2, 1 & - \\
%&  0,1,-1,-1,7,2,1 & + \\
8_2 & 0, -1, 1, -1, -1, 0, -1 & + \\
8_3 & 0, 1, 1, 1, 5, 0, -1 & - \\
8_4 & 0, 1, 3, -1, 1, 2, -1 & - \\
8_5 & 0, 1, -1, -1, 3, 0, -1 & - \\
8_6 & 0, -1, 3, -1, -1, 0, 1 & + \\
8_7 & 0, -1, 1, 1, -1, 0, 1 & + \\
8_8 & 0, 1, 1, 1, -1, 0, 3 & + \\
8_9 & ??? & \\
8_{10} &   0,1,-1,-1,-1,0,3 & +  \\
8_{11} & 0, 3, 1, -1, -1, 0, 1 & - \\
%  &  0,1,-1,1,5,4,-1 & + \\
8_{12} & 0, 1, 1, -1, 1, 2, -1 & + \\
8_{13} & 0, 1, -1, 3, -1, 2, 1 & + \\
8_{14} & 0, -1, -1, -1, 1, 2, -1 & + \\
8_{15} & 0, 1, 1, -1, -1, -2, -1 & - \\
8_{16} & -2, 1, -1, 1, 3, 0, -3 & + \\
%   & 0, 1, 1, 3, 1, 0, -3 - \\
8_{17} & 0, 1, 1, -1, 1, 0, 1 & + \\
8_{18} & ??? & \\
8_{19} & 0, 1, -1, -1, -1, 0, -1 & - \\
8_{20} & 0, 1, -1, 1, -1, 0, 1 & + \\
8_{21} & 0, 1, 1, -1, 1, 0, -1 & - \\
\hline
\end{array}
$$

$$
\begin{array}{c|c|c}
{\rm knot} &   n_1,n_2,n_3,n_4, n_5,n_6,n_7 & \pm \\
 \hline
9_1 & ??? \\
9_2 & 2, -1, -1, -1, -1, -2, 1 & + \\
9_3 & 0, -1, 1, 1, -3, 0, -3 & - \\
9_4 & 0, -1, -1, -1, -5, 2, 1 & + \\
%  & 0, -1, -1, -1, -1, 2, 1 & - \\
9_5 & 0, 1, -1, -1, -3, 2, 1 & - \\
9_6 & -2, -1, 1, -1, -1, 0, -1 & - \\
9_7 & 0, -1, -1, -1, -1, -2, 1 & + \\
9_8 & 0, 1, -1, -1, 3, 2, 3 & + \\
9_9 & 0, -1, -1, -1, -1, 0, 1 & - \\
9_{10} & 0, -1, 1, 1, -3, 0, -1 & - \\
9_{11} & 0, -1, 1, 1, -1, 0, 1 & - \\
9_{12} & 0, 1, -1, 1, 1, -2, 1 & - \\
%  & ??? 0,1,-1,7,1,4,-3 & + \\
9_{13} & 0, 1, -1, -1, -3, 0, 1 & - \\
9_{14} & 0, 1, 3, 3, 1, 2, -1 & - \\
%  &  0,1,-1,1,-3,4,-1 & + \\
9_{15} & 0, 1, -1, 1, -1, -2, -1 & + \\
9_{16} & -2, -1, -1, 1, -1, -2, 1 & + \\
9_{17} & 0, 3, 1, 1, 3, -2, -1 & - \\
9_{18} & 0, 1, -3, -1, -1, 0, 1 & -\\
9_{19} & 0, 1, -1, -1, 3, 2, -1 & + \\
9_{20} & 0, -1, 1, -1, -1, -2, 1 & + \\
9_{21} & 0, 1, -1, 3, 3, 2, 1 & + \\
9_{22} & 0, 1, -1, -1, 3, 0, -1 & + \\
9_{23} & 0, 1, -1, -1, -1, -2, 1 & - \\
9_{24} & 0, 1, -1, -1, 3, 0, 3 & + \\
9_{25} & 0, 1, 1, -1, -1, -2, -1 & + \\
9_{26} & 0, -1, 1, -1, -1, 2, 1 & + \\
9_{27} & 0, -1, 1, 1, 3, 0, 1 & + \\
9_{28} & ??? & \\
9_{29} & 0, 1, -1, 3, -1, 0, -1 + \\
9_{30} & 0, -1, 3, 1, 1, 0, 1 & + \\
9_{31} & -2, 1, -1, -1, 1, 0, 1 & + \\
9_{32} & 0, 1, 1, -1, 1, 0, 1 & - \\
%  &  2,1,1,-1,-3,0,-1 & + \\
9_{33} & 0, 1, -3, -1, 1, 4, 1 & + \\
9_{34} & 2, 1, 3, 1, -1, 0, 1 & - \\
9_{35} & 2, -3, -1, -1, -3, 2, 1 & + \\
%  & 2,-1,1,1,-3,0,-1 & - \\
9_{36} & 0, 1, -1, -1, -1, 0, 3 & - \\
%  &  2,1,-3,1,-1,-2,1 & + \\
9_{37} & 0, -1, 3, 1, 1, 2, -1 & + \\
9_{38} & 0, 1, -3, -1, -1, 2, -1 & - \\
9_{39} & 2, -1, -1, 1, -1, 0, -1 & + \\
9_{40} & ??? & \\
9_{41} & 0,1,1,-1,-3,2,1 & + \\
9_{42} & 0, 1, 1, -1, 1, 0, -1 & + \\
9_{43} & 0, -1, -1, -1, 1, 0, -1 & + \\
9_{44} & 0, -1, 1, 1, 1, 2, 1 & + \\
9_{45} & 0, 1, -1, 1, -1, 0, 1 & - \\
%   &   0,1,-1,-3,-1,0,1 & + \\
9_{46} & 2, 3, -1, -1, 3, 2, 1 & + \\
9_{47} & 0, -1, 3, 1, -1, 0, -1 & + \\
9_{48} & 0, 1, -3, -1, -1, 2, 1 & + \\
9_{49} & 0, 1, 1, -1, -3, 0, -1 & - \\
\end{array}
$$

$$
\begin{array}{c|c|c}
{\rm knot} &   n_1,n_2,n_3,n_4, n_5,n_6,n_7 & \pm \\
 \hline
10_{1} & 0, 1, -1, 1, 1, 10, -1 & +  \\
% &  0, 1, -1, 3, 1, 2, 5 & - \\
10_{2} & ??? &  \\
10_{3} & 0, 1, -1, -1, 5, 2, 1 &  - \\
10_{4} & 0, 1, 5, -1, 1, 2, -1 & -  \\
10_{5} & ??? &  \\
10_{6} & 0, -1, 1, -1, -1, 0, -3 & + \\
10_{7} &  0,1,-1,1,7,4,-1 & + \\
10_{8} & 0, -1, 1, 1, 5, -2, -1 & - \\
10_{9} & ??? &  \\
10_{10} &  0,1,-1,1,-5,2,1 &  + \\
10_{11} & 0, 1, -1, -1, 5, 0, 1 &  - \\
10_{12} & 0, -1, -1, -1, -1, 2, 3 & + \\
10_{13} & 0, 1, 3, -1, 1, 2, -1 & + \\
10_{14} & 0, -1, -1, -1, -1, 2, -1 & + \\
10_{15} & 0, -1, 1, 1, -1, 0, 3 & + \\
10_{16} & 0, 1, -3, 1, 5, 0, 3 &  - \\
10_{17} & ??? &  \\
10_{18} & 2, -1, 3, -1, -1, -2, 1 & + \\
10_{19} & 0, -1, 1, 1, -3, 4, 1 & + \\
10_{20} &  0,1,1,1,3,6,-1 & +  \\
10_{21} & 0, -1, 1, 1, -3, 0, -3 &  + \\
10_{22} & ??? &  \\
10_{23} &  0,-3,1,3,-1,0,-1 & + \\
10_{24} & 0,3,-3,1,1,0,-1 & - \\
% & 0,1,-1,1,5,6,-1&  + \\
10_{25} & 0, 1, -1, -1, 3, -2, -1 & -  \\
10_{26} & ??? &  \\
10_{27} & 0,1,1,-1,-5,0,1 & + \\
10_{28} & 0, 1, -3, -3, 1, 2, 3 & + \\
10_{29} & ??? &  \\
10_{30} & 2, -1, -1, -1, -1, 2, -1 & + \\
10_{31} & 2, -1, 1, 1, -1, 0, 3 & + \\
10_{32} & -2, 1, 1, -1, 3, 0, 1 & + \\
10_{33} &  0,1,-1,5,-1,2,-1 &  + \\
10_{34} & 2, -1, -1, 3, 1, 2, 3 & + \\
10_{35} & 0, 1, -1, -1, 3, 2, 3 & - \\
10_{36} & 0, -1, -1, -1, 3, -2, 1 & + \\
10_{37} & 0, -1, -1, 1, 1, 2, 3 & + \\
10_{38} & 0, 1, -1, -1, 3, -2, 1 & - \\
10_{39} & -2, -1, 1, -1, -1, 0, -1 & + \\
10_{40} & -2,-1,1,1,-1,0,1 & + \\
10_{41} & 0, 3, 1, -1, -1, 2, 1 & - \\
%  &  2,1,3,1,1,0,-1 & + \\
10_{42} & -2, 1, 3, 1, 1, 0, -1 & + \\
10_{43} & -2, 1, -1, -1, 1, 2, 3 & + \\
10_{44} & 0, -1, 1, 1, 3, -2, -1 & + \\
10_{45} & 0, -1, 1, 1, 3, 2, -1 & + \\
10_{46} & ??? &  \\
10_{47} & ??? &  \\
10_{48} & ??? &  \\
10_{49} & ??? &  \\
10_{50} & 0, 1, -1, -3, 3, 0, -1 & - \\
10_{51} & 0, 1, -3, -1, -1, 0, 3 & + \\
10_{52} & 2, -1, 1, 3, 1, -2, 1 & + \\
10_{53} & 0, 1, -3, -3, 1, 0, -1 & - \\
10_{54} & 0, -1, -1, 1, 1, 0, 3 & + \\
10_{55} & 2, 1, -1, -1, -1, -2, -1 & - \\
%  & 0, 1, -3, -1, 1, -2, -1 & -
\end{array}
$$

$$
\begin{array}{c|c|c}
{\rm knot} &   n_1,n_2,n_3,n_4, n_5,n_6,n_7 & \pm \\
 \hline
10_{56} &  0, 1, -1, -1, 3, -2, -1 & - \\
10_{57} & 0, -1, -1, -1, 1, -2, 3 & + \\
10_{58} & 0, 1, 1, 1, 3, 2, 1 & - \\
%  &  0,1,-1,-3,3,4,1 & + \\
10_{59} & 0, 1, 3, -1, -1, 0, 3 - \\
10_{60} & 2, 1, 1, -1, -1, 2, 3 & - \\
10_{61} & 0, 1, -1, -1, 5, 0, -1 & -  \\
10_{62} & 0, -1, -1, -1, -1, 0, 3 & + \\
10_{63} & 0, 1, -3, -3, 1, 2, -3 &  - \\
10_{64} & ??? &  \\
10_{65} & 2, -1, -1, -1, -1, 0, 3 & + \\
10_{66} & 0, -1, 1, -1, -1, -2, -1 & - \\
10_{67} & 2, 1, 1, 1, 3, 0, 1 & + \\
10_{68} &  2,-3,1,3,-1,0,-1 & + \\
10_{69} &  2,-3,-1,-3,1,0,3 & + \\
10_{70} & 0, 1, -1, -1, 3, 0, 3 & - \\
10_{71} & -2, -1, 1, 1, 1, 2, 1 & + \\
10_{72} & -2, -1, -1, -1, 1, 0, -1 & + \\
10_{73} & 2, 1, 1, -1, -1, 2, -1 & -  \\
10_{74} & 2,3,1,1,3,0,-1 & + \\
10_{75} & 0, 3, 1, 1, 3, 0, -3 & +  \\
10_{76} & ??? &  \\
10_{77} & -2, -1, -1, -1, 1, 0, 3 & + \\
10_{78} & 2, 1, 1, -1, -1, -2, -1 & -  \\
10_{79} & ??? &  \\
10_{80} & ??? &  \\
10_{81} & 2, -1, -1, 1, 3, -2, 3 & + \\
10_{82} & ??? &  \\
10_{83} & 2, -1, 1, 3, 1, 2, 1 & + \\
10_{84} & -2, -1, -1, 1, 3, 0, -1 & + \\
10_{85} & ??? &  \\
10_{86} & 2, -1, 3, 1, 1, -2, 1 & + \\
10_{87} & 0, 1, 1, -1, 1, -2, 1 & + \\
10_{88} & 2, -1, -1, 1, 3, 2, -1 & + \\
10_{89} & 2,1,-1,-3,-1,0,1 & + \\
10_{90} & -2, 1, 3, -1, 1, 2, 1 & + \\
10_{91} & ??? &  \\
10_{92} & 0, 1, 3, -1, -1, -2, -1 & - \\
10_{93} & 0, 1, 1, 5, 1, 0, -3 & - \\
10_{94} & ??? &  \\
10_{95} &  0,1,-1,-3,-1,0,3 & + \\
10_{96} & ??? &  \\
10_{97} & 0, 1, -3, -1, -1, 2, -1 & + \\
10_{98} & 2, 3, -1, -1, -1, 0, -1 & - \\
10_{99} & ??? &  \\
10_{100} & ??? &  \\
10_{101} & 0, 1, 1, -1, -3, -2, -1 & -  \\
10_{102} & 0, 1, 1, -1, 3, 0, 1 & + \\
10_{103} & 0,-1,-3,1,3,0,-1 &  + \\
10_{104} & ??? &  \\
10_{105} & -2, 1, 1, -1, 1, 0, -1 & + \\
10_{106} & 0, -1, 1, -1, 1, 0, 1 & + \\
10_{107} & 0, -1, -1, 1, 3, 2, 1 & + \\
10_{108} & 0, 1, 1, 3, 3, 0, -3 & -  \\
10_{109} & ??? &  \\
10_{110} & 0, -1, 1, -1, 1, 2, -1 & +
\end{array}
$$

$$
\begin{array}{c|c|c}
{\rm knot} &   n_1,n_2,n_3,n_4, n_5,n_6,n_7 & \pm \\
 \hline
10_{111} & 0, 1, -3, -1, 3, 0, -1 & - \\
10_{112} & ??? &  \\
10_{113} & -2, -1, 3, 1, -1, 0, -1 & + \\
10_{114} & ??? &  \\
10_{115} & ??? &  \\
10_{116} & ??? &  \\
10_{117} & 0, 1, 3, 3, -1, -2, -1 & - \\
%&  -2,1,-3,1,3,0,-3 & + \\
10_{118} & ??? &  \\
10_{119} & ??? &  \\
10_{120} & ??? &  \\
10_{121} & ??? &  \\
10_{122} & ??? &  \\
10_{123} & ??? &  \\
10_{124} & ??? &  \\
10_{125} & ??? &  \\
10_{126} & ??? &  \\
10_{127} & ??? &  \\
10_{128} & 0, 1, -3, -1, -1, 0, -1 & - \\
10_{129} & 0, 1, -3, 5, 1, 0, -1 & + \\
10_{130} & 0, 1, -1, 3, 3, 0, 1 & + \\
10_{131} & 0, 1, 1, -3, 1, 0, -1 & -  \\
10_{132} & 0,1,1,1,-1,-4,1  & + \\
%& 0, 1, -3, -1, 1, 2, -1 & - \\
10_{133} & 0, 1, 1, -1, 1, -2, -1 & - \\
10_{134} & 0, 1, -1, -1, -1, -2, -1 & - \\
10_{135} & 0, 1, -1, 1, -1, 0, 3 & + \\
10_{136} & 0, 1, 1, -1, -1, 2, 3 & - \\
% &  0,1,-1,-3,-1,4,1 & + \\
10_{137} & 0, 1, 1, -1, -1, 2, -1 & + \\
10_{138} & 0, 1, 3, -1, -1, 0, -1 & + \\
10_{139} & 0, -1, -1, -1, -1, 0, -1 & -  \\
10_{140} & 2, -1, 1, -1, -1, -2, 3 & + \\
10_{141} & ??? &  \\
10_{142} & 0, 1, -1, -1, -3, 0, -1 & -  \\
10_{143} & 0, -1, 1, 1, -1, -2, 1& + \\
10_{144} & 0, 1, 3, -1, 1, 0, -1 & - \\
10_{145} & 0, 1, -3, -1, -1, -2, 1 & +  \\
10_{146} & 0, 1, -1, 3, -1, 2, -1 & + \\
10_{147} & 0, -1, 3, 1, 1, -2, -1 & + \\
10_{148} & 0, -1, -1, 1, -1, 0, 1 & + \\
10_{149} & 0, -1, 1, -1, -3, 0, -1 & + \\
10_{150} & 2, 1, -1, 1, 3, 0, 1 & + \\
10_{151} & 0, -1, -1, 1, -1, 2, 1 & + \\
10_{152} & ??? \\
10_{153} & ??? \\
10_{154} & 0, 1, -3, -1, 1, 0, -3 & - \\
10_{155} & 0, -1, 1, -1, 1, 2, 1 & + \\
10_{156} & 0, 1, 1, 1, -1, 0, 3 & + \\
10_{157} & 0, -1, 3, 1, -1, -2, -1 & - \\
10_{158} & 0, -1, 3, 1, 3, 0, -1 & + \\
10_{159} & -2, 1, 1, -1, -3, 0, 1 & + \\
10_{160} & 0, -1, -1, 1, -1, 0, -1 & + \\
10_{161} & 0, 1, 1, -1, -3, 0, -3 & - \\
10_{162} & 0, 1, 1, -1, 3, 0, -1 & - \\
10_{163} & ??? &  \\
10_{164} & 2, -1, 1, 3, 1, 0, -1 & + \\
10_{165} & 2,-1,-3,1,3,0,-1 & - \\
\end{array}
$$

\noindent
In particular, for twist knots $H^{tw(k)}= H^{(0, 1, -1, 1, 1, 2-2k, -1|\pm)}$
irrespectively of the sign at the last position.
We remind that $3_1=tw(1)$, while $4_1 = tw(-1)$, and in general
$(2k+1)_2=tw(k)$, while $(2k+2)_1=tw(-k)$.

\bigskip

\noindent
Note that, because of additional powers of $T$-matrices in $K$ and $\bar K$,
one can not just invert the signs of all $n_i$ in the lines of the above table.
For the same reason, one should not be surprised if the sum of $n_i$ in the
table is smaller than the intersection number of the knot: there are still
six additional crossings hidden in $K$ and $\bar K$.
Unfortunately, this makes above identification of knots not fully reliable:
it can happen that some entries in the table actually describe knots with
eleven or more intersections.
This is easy to check by comparison with Jones polynomials in representation $[2]$,
but we did not perform this check for the whole list.
Thus the data in the table should be taken with a certain care.

Moreover, there are five pairs of knots with even {\it less} than $11$ intersections:
$5_1 \& 10_{132}$,  $8_8\&10_{129}$,
$8_{16}\&10_{156}$, $10_{25}\& 10_{56}$, $10_{40}\&10_{103}$,
which are not distinguished by the fundamental HOMFLY polynomials.
To separate them we {\it did} look at Jones$_{[2]}$.

Starting from $11$ crossings there will be pairs of mutants,
which are not distinguished by {\it any} symmetrically colored HOMFLY polynomial.
Exactly at eleven intersections there are $16$ such pairs,
somewhat mysteriously no one showed up in our analysis of the family
(\ref{7parfam}).
Moreover, beginning from $16$ intersections there are mutants
inseparable even by $H_{[21]}$, see \cite{Morton} and \cite{MMMRS}.

\end{document}